# Distributed Snitch Digital Twin-Based Anomaly Detection for Smart Voltage Source Converter-Enabled Wind Power Systems

Mohammad Ashraf Hossain Sadi, *Senior Member, IEEE,* Soham Ghosh, *Senior Member, IEEE,* Siby Plathottam, *Member, IEEE,* Mohd. Hasan Ali, *Senior Member, IEEE*

*Abstract*—Existing cyberattack detection methods for smart grids such as Artificial Neural Networks (ANNs) and Deep Reinforcement Learning (DRL) often suffer from limited adaptability, delayed response, and inadequate coordination in distributed energy systems. These techniques may struggle to detect stealthy or coordinated attacks, especially under communication delays or system uncertainties. This paper proposes a novel Snitch Digital Twin (Snitch-DT) architecture for cyber-physical anomaly detection in grid-connected wind farms using Smart Voltage Source Converters (VSCs). Each wind generator is equipped with a local Snitch-DT that compares real-time operational data with high-fidelity digital models and generates trust scores for measured signals. These trust scores are coordinated across nodes to detect distributed or stealthy cyberattacks. The performance of the Snitch-DT system is benchmarked against previously published Artificial Neural Network (ANN) and Deep Reinforcement Learning (DRL)-based detection frameworks. Simulation results using an IEEE 39-bus wind-integrated test system demonstrate improved attack detection accuracy, faster response time, and higher robustness under various cyberattack scenarios.

## I. Introduction

To achieve the goal of 100% clean energy by 2035 [1], the deployment of smart Voltage Source Converter (VSC)-enabled Distributed Energy Resources (DERs), such as wind generators, plays a pivotal role in enhancing the resilience, reliability, and flexibility of modern power systems. These DERs, characterized by decentralized ownership, diverse operational conditions, and dynamic control features, introduce new challenges in coordination and cybersecurity. The Smart Inverter (SI), a firmware-driven embedded VSC system, performs traditional grid interfacing functions (DC/AC and AC/DC conversion) while also supporting high-speed communication, remote control, edge/cloud computing, and advanced control schemes [2].

The growing adoption of smart VSCs in wind farms opens promising avenues for grid-supportive functionalities including autonomous voltage support, Volt-Var (VV) and Volt-Watt (VW) control, and real-time control orchestration across distributed assets. Standards such as IEEE 2800-2022 and IEEE 1547 [3],[4] have established interoperability benchmarks for these inverter-based resources (IBRs), enabling them to actively participate in voltage and reactive power regulation. Unlike legacy systems where capacitor banks and tap changers dominated VV/VW control, smart VSCs incorporate these functions natively into their control loops, making coordinated operation with legacy systems more critical than ever [5].

However, the increasing digitization and communication capabilities of these smart VSCs have also made them attractive targets for cyber adversaries. Despite standard compliance, attackers can exploit setpoint vulnerabilities in VV and VW control functions to introduce False Data Injection (FDI) attacks, leading to abnormal responses such as voltage oscillations, equipment damage, and cascading system failures.

As reported by the Department of Energy (DOE), cyber intrusion events in the energy sector have been consistently rising, with 36 incidents recorded in early 2023 alone [6]. Several recent studies have attempted to address these security concerns through attack detection and mitigation mechanisms. Long Short-Term Memory (LSTM) and other neural network-based approaches have been employed to detect anomalies in photovoltaic systems and microgrids [7], [8]. Supervised learning techniques have been used to track reactive power setpoints and mitigate false data [9], while Deep Reinforcement Learning (DRL) methods such as Proximal Policy Optimization (PPO) have demonstrated potential in learning adaptive control policies for cyberattack mitigation [10], [11]. Blockchain-enabled firmware verification modules [12] and Hardware Security Modules (HSMs) [13] have also been proposed to reinforce the physical security perimeter of inverter firmware.

Despite these advancements, most of the existing solutions either focus on centralized detection strategies or require extensive offline training. They often lack real-time explainability limiting their applicability in large-scale, wind-integrated grid environments. Additionally, most detection and mitigation approaches treat each inverter as an isolated unit, which overlooks the spatially correlated nature of many cyberattacks in interconnected DER networks [14], [15].

Recent advances in digital twin (DT) technologies have enabled real-time, high-fidelity modeling of power electronic systems for enhanced monitoring, control, and cybersecurity. A particularly promising variant is the Snitch Digital Twin approach, which extends traditional DTs by incorporating anomaly detection logic and trust evaluation mechanisms into the digital replica. Originally proposed in the context of cyber-physical systems for industrial and energy applications, the Snitch-DT acts as a silent observer that "snitches" on malicious or abnormal behaviors by comparing real-time physical data with simulated outputs from a synchronized virtual model.

To address these limitations, this paper proposes a novel Snitch Digital Twin (Snitch-DT) framework for anomaly detection in smart VSC-enabled wind power systems. In the proposed architecture, each wind generator is equipped with a local Snitch-DT module that mirrors the real-time dynamics of the physical VSC system. These twins compute residuals

Dr. M.A.H. Sadi is with the University of Central Missouri, MO, USA (sadi@ucmo.edu). Soham Ghosh is with Black & Veatch, Kansas, 66211, USA. Dr. Siby Plathottam is with Argonne National Laboratory, Lemont, IL, USA. Dr. M.H. Ali is with the University of Memphis, Memphis, TN, USA.

between simulated and measured system variables and generate time evolving trust scores for each control input and sensor stream. By enabling coordination across multiple Snitch-DT instances, the system can detect not only isolated anomalies but also stealthy and coordinated attacks that span across multiple DER nodes.

The objective of this work is to design and validate a cooperative, and real-time anomaly detection framework using Snitch Digital Twins across smart VSC-equipped wind generators. The detection performance of the proposed Snitch-DT architecture is then benchmarked against the authors previously developed Artificial Neural Network (ANN) [16] and Deep Reinforcement Learning (DRL) based [17] detection techniques using the same grid structure, attack scenarios, and system metrics.

The key contributions of this paper are summarized as follows. First, a high-fidelity Snitch-DT model is developed for each inverter node in a multi-wind-generator system, enabling local anomaly detection through residual and trust score computation. Second, a distributed coordination mechanism is introduced that aggregates trust scores to improve attack detection robustness against spatially correlated attacks. Third, the system is tested under various FDI attack types including bias, ramp, and delay manipulations using a detailed MATLAB/Simulink model of the IEEE 39-bus test system with wind integration [17] as represented in Fig.1. Finally, the detection accuracy, latency, and robustness of the Snitch-DT architecture are compared against DRL and ANN based mitigation strategies under identical simulation conditions.

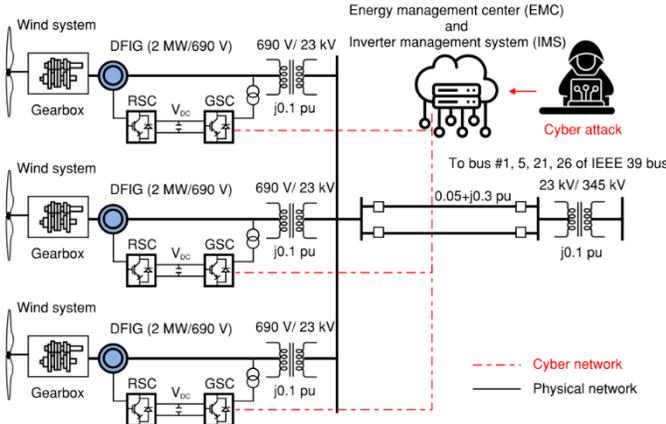

Fig. 1. Cyber physical structure of the wind farm connected in bus 1 of the IEEE 39 bus test system. Wind farms are also connected in bus 5, 21, and 26 of the IEEE 39 bus system.

## II. CYBERATTACK MODELING AND DETECTION

In the control architecture of a doubly fed induction generator (DFIG)-based wind generator, the Grid Side Converter (GSC) is responsible for reactive power compensation to maintain voltage regulation and ensure grid stability [18]. Under normal operation, the GSC regulates the grid reactive power, $Q_g$ according to the voltage-reactive power (Q–V) droop characteristic defined by the secondary PI controller. However, malicious False Data Injection (FDI) attacks targeting the GSC's reactive power setpoints can cause abnormal power flows, voltage instability, and long-term damage to inverter assets. In this work, we propose a novel anomaly detection method using a Snitch-DT framework to detect such attacks in real time.

### A. Digital Twin-Based Cyberattack Detection Framework

The Snitch-DT module is a high-fidelity digital replica [19] of the physical smart VSC controller and is co-simulated in real time with the physical system. Unlike traditional state estimators that infer internal system states based solely on real-time measurements and system models, Digital Twins (DTs) maintain a continuously running, physics-informed replica of the actual system, enabling the simulation of system dynamics beyond just current states. As illustrated in Fig. 2, for each wind generator node, the Snitch-DT continuously receives the input signals from the physical inverter system, such as the grid voltage $V_g(t)$, grid current $I_g(t)$, and measured grid reactive power $Q_g(t)$. Digital Twin (DT) also simulates its own estimate of these values $\widehat{V^*}_{ot}(t)$ using the same control model and known references under the assumption of no attack. Any discrepancy between the simulated and real system output forms the residual signal:

$$r(t) = V_{ot}(t) - \widehat{V^*}_{ot}(t) \qquad (1)$$

This residual is then passed through a statistical anomaly thresholding mechanism to identify abnormal behavior. The detection condition can be defined as:

$$\textit{Anomaly detected if } |r(t)| > \epsilon \qquad (2)$$

where $\epsilon$ is a calibrated threshold based on system noise and modeling error. In addition to residual generation, each Snitch-DT computes a dynamic trust score for the monitored GSC, quantifying the probability of compromise. The trust score $\tau(t) \in [0,1]$ is updated over time using a decaying window of past residuals:

$$\tau(t) = \exp\left(-\frac{1}{\Delta T} \sum_{k=t-\Delta T}^{t} \frac{r^2(k)}{\sigma^2}\right) \qquad (3)$$

where $\Delta T$ is the rolling time window and $\sigma^2$ is the variance of residuals under healthy operation.

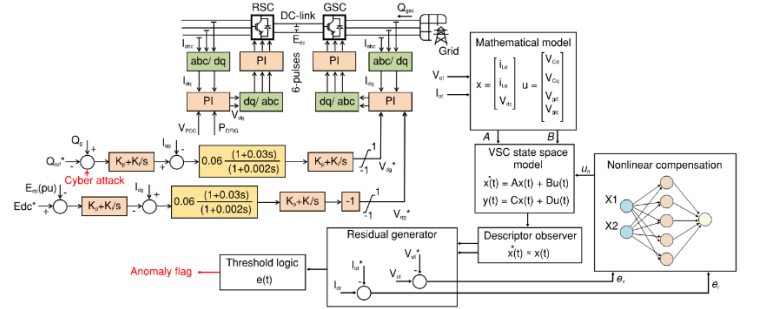

Fig.2. Proposed control architecture to detect the cyberattack in the voltage source converter.

Unlike centralized or node-isolated anomaly detection methods, the proposed framework integrates multiple Snitch-DT instances one per wind generator to form a cooperative detection network. Each node communicates its trust score to its peers, and a consensus mechanism aggregates the scores to assess whether the system is under attack locally or globally.

This helps detect coordinated multi-node attacks, which are often missed in localized schemes.

### B. Deep Reinforcement Learning (DRL) Based Detection

The Deep Reinforcement Learning (DRL) method, presented in our earlier work [17], uses a Deep Q-Network (DQN) framework to detect anomalies in smart VSC control loops of wind generators under cyberattack. The problem is modeled as a Markov Decision Process (MDP), where observed states—comprising terminal voltage and output current—serve as input to a neural network-based Q-function.

The agent learns an optimal policy offline using historical data by minimizing the temporal-difference (TD) loss. The Q-network, with two ReLU-activated hidden layers, estimates Q-values for possible actions. Once trained, the agent identifies abnormal patterns in control signals by tracking deviations from reference trajectories and signaling anomalies based on learned behavior.

### C. Supervised Learning Based Detection

As a benchmark, a supervised Artificial Neural Network (ANN) model developed in our previous work [16] is used to detect False Data Injection (FDI) attacks targeting the reactive power control loop of Smart VSCs.

The ANN is trained to learn the nonlinear relationship between grid-side voltage $V_g(t)$ and current $I_g(t)$ over a fixed observation window. With $N_i = N_r \times N_m$ input features, the network outputs a predicted reactive power setpoint $\hat{Q}_g^*$. During online operation, the deviation $|Q_g(t) - \hat{Q}_g^*|$ is monitored, and an alarm is triggered if it exceeds a set threshold $\epsilon$. This lightweight method is deployed locally at inverter nodes for scalable, real-time detection.

## III. SIMULATIONS AND ANALYSIS

### A. Simulation Setup

To evaluate the performance of the proposed Snitch Digital Twin (Snitch-DT)-based anomaly detection framework, simulations are conducted on a modified IEEE 39-bus system in MATLAB/Simulink. Wind generator models are integrated at buses 1, 5, 21, and 26, each interfaced with the grid through smart VSCs. The wind generators and loads are scaled appropriately to conform to the IEEE 39-bus operational limits. The Snitch-DT framework is deployed locally at each wind generator, operating in parallel for distributed detection.

Each VSC is modeled to support Volt-Var (VV) and Volt-Watt (VW) control functionalities, in line with IEEE 1547 and IEEE 2800 standards [3],[4]. Detection logic is implemented through user-defined function blocks that interface with Simulink through embedded MATLAB scripts. The digital twin models replicate VSC behavior based on dynamic system parameters and serve as real-time reference systems for anomaly detection via residual analysis.

Three detection strategies are implemented independently for comparison:
- Proposed Snitch-DT detection (model-based residual monitoring),
- Supervised learning-based ANN detection (data-driven statistical estimation),
- DRL-Based detection using PPO Agent (policy-based anomaly recognition).

To validate detection effectiveness under diverse threat conditions, four distinct cyberattack scenarios are simulated: No Attack, False Data Injection (FDI) for bias and ramp attack, delay attack, and coordinated cross-node attack. For each attack type 10 different unique scenarios are considered.

For each scenario, system performance is assessed based on anomaly detection accuracy, False Positive Rate (FPR), False Negative Rate (FNR), detection delay, Root Mean Square Error (RMSE) in reactive power tracking, and computation time [16]. Simulation time steps are fixed to ensure deterministic evaluation across sessions.

### B. Case I: False Data Injection (FDI)-Bias and Ramp Attack

In this scenario, two types of data manipulation attacks are simulated: a bias injection and a ramp attack, both targeting the reactive power setpoint $Q_g^*$ of the wind generator's grid-side voltage source converter (GSC). For the bias FDI attack, a step offset of +0.1 pu is injected at t=0.1 sec. This results in a consistent upward deviation in the reactive power reference, forcing the VSC to supply more reactive power than required. The ramp FDI attack, also triggered at t=0.1 sec, introduces a gradually increasing positive ramp over time, mimicking stealthy or progressive attacks. Fig. 3 and Fig. 4 represent the reactive power and voltage response of Wind Generator 1 under bias and ramp attacks respectively for three different detection cases. Without any detection mechanism, both bias and ramp attacks result in significant deviations in the reactive power output, destabilizing voltage regulation and potentially damaging connected equipment. With the implementation of the Snitch-DT framework, the attack-induced deviation is swiftly identified by comparing real and virtual measurements. The Snitch-DT offers superior tracking accuracy and fast stabilization compared to the other methods. This happens due to its ability to override the discrepancy with the digital twin derived estimations. The DRL-based detection approach, though more adaptive than the supervised ANN method, presents moderately higher overshoot and a longer settling time due to the inherent exploration-exploitation balance in policy learning. The ANN-based detection method performs the weakest among the three, showing slower detection latency and limited generalization.

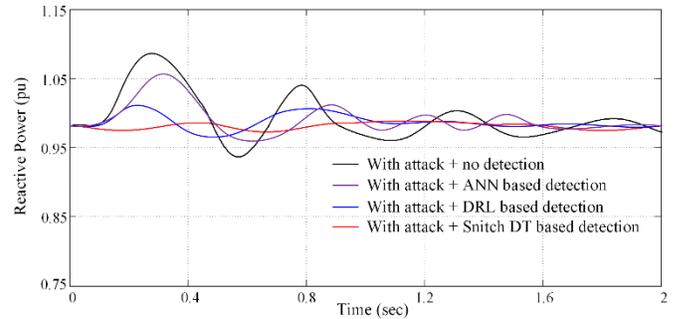

Fig.3. Reactive power response of the wind generator 1 for 0.1 pu bias attack.

### C. Case II: Delay Attack

In this scenario, a data latency cyberattack is introduced at t = 0.2 s by injecting a 20 ms time delay into the feedback path of the smart Voltage Source Converter (VSC) controller in the

Simulink environment. This results in the VSC operating based on outdated grid measurements, leading to degraded control performance. As shown in Fig. 5, the terminal voltage of the wind generator experiences a noticeable deviation due to delayed control response. The voltage rises above the nominal 1 pu value and exhibits increased settling time and instability. Among the compared detection methods, the Snitch Digital Twin (Snitch-DT) framework demonstrates the fastest recognition and isolation of the latency-induced anomaly, effectively restoring voltage to nominal levels. In contrast, the DRL-based method provides moderate correction with slight delay, while the ANN-based detection lags further in response, allowing more pronounced voltage drift.

### D. Anomaly Detection Performance Evaluation

To evaluate and compare the anomaly detection performance of the considered methods — Snitch-DT, DRL, and ANN, multiple performance metrics are analyzed: accuracy, FPR, FNR, detection delay, RMSE in reactive power tracking, computation time, and F1 score. These metrics offer a holistic view of both detection efficacy and real-time feasibility.

In this context, False Positives (FP) refer to benign operational states misclassified as anomalies, while False Negatives (FN) correspond to actual attack instances failing to be detected. While both are undesirable, a high FN rate is particularly critical as it implies missed cyberattacks. Detection accuracy, precision, and recall are calculated as:

$$\left. \begin{array}{l} Accuracy = \frac{No. of\ (TP+TN)}{No. of\ (FP+FN+TP+TN)} \\ Precision = \frac{No. of\ TP}{No. of\ (FP+TP)} \\ Recall = \frac{No. of\ TP}{No. of\ (FN+TP)} \end{array} \right\} \quad (4)$$

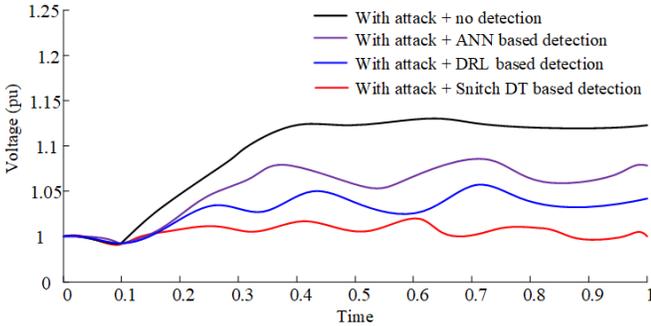

Fig.4. Voltage response of the wind generator 1 for 0.1 pu ramp attack at 0.1 sec.

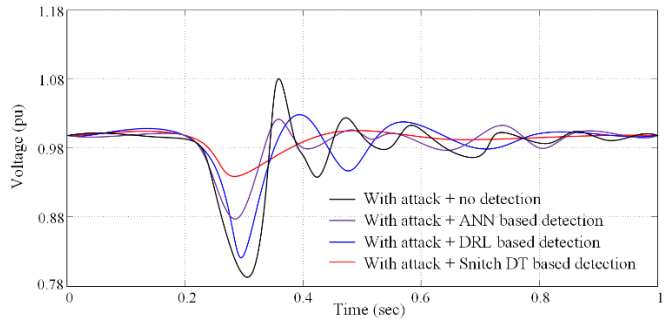

Fig.5. Voltage response of the wind generator 1 for 20 ms delay attack at 0.2 sec.

Based on Precision and Recall, we incorporate the F1 score as a comprehensive evaluation metric, which balances both precision and recall, and offers a more reliable indication of detection performance under diverse attack scenarios.

$$F1\ Score\ =\ 2 * \frac{Precision * Recall}{Precision + Recall} \quad (5)$$

From the comparative bar chart in Fig. 6, Snitch-DT exhibits superior performance with the highest accuracy of 95%, lowest FPR (10%), and lowest FNR (8%) among all methods. In terms of detection delay, Snitch-DT also achieves the fastest response (100 steps) compared to 400 steps for DRL and 600 steps for ANN. Its RMSE in tracking the reactive power reference is only 0.05 pu, confirming tight precision, whereas DRL and ANN show 0.15 pu and 0.25 pu, respectively. Here the RMSE indicates the reactive power tracking accuracy post-detection. Computation time also favors Snitch-DT, making it more viable for real-time anomaly detection applications in VSC systems.

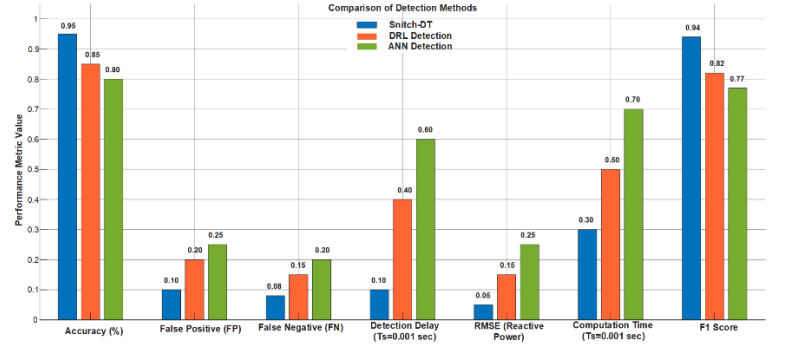

Fig. 6. Comparison of the three detection methods in terms of different indices.

Further, the Receiver Operating Characteristic (ROC) curves shown in Fig. 7 reinforce these findings. The Snitch-DT curve remains consistently above those of DRL and ANN, achieving the highest Area Under Curve (AUC) score, indicating strong discriminative power and low generalization error. While DRL follows closely with moderate AUC and slope, ANN shows more false positives across the spectrum, especially at lower detection thresholds.

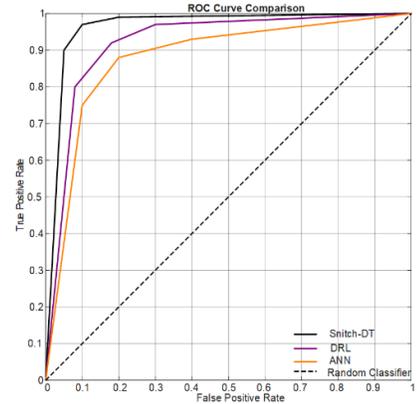

Fig.7. ROC curve comparison between three detection methods.

These observations confirm that Snitch-DT not only detects anomalies faster and more accurately but also ensures smoother system behavior during and after attacks. Its localized modeling through real-time mirrored virtual twins enables proactive inference of inconsistencies in the control layer, providing an advantage over policy-based DRL and purely data-driven ANN classifiers.

Overall, the performance analysis confirms that Snitch-DT achieves a balanced trade-off between speed, precision, and computational overhead, outperforming both DRL and ANN-based techniques across diverse cyberattack scenarios. A comprehensive comparison among all the three considered cyberattack detection methods is presented in Table I.

TABLE I
COMPARISON AMONG DIFFERENT DETECTION METHODS

|  | Snitch Digital Twin (proposed) | DRL based detection | Supervised ANN based detection |
|---|---|---|---|
| Approach type | Physics-informed + Residual-based digital twin model | Model-free reinforcement learning using PPO agent | Supervised learning with trained feedforward ANN |
| Detection Logic | Residual generation with statistical thresholding | Reward signal deviation; policy-based detection using critic evaluation | Compare real to learned reference via ANN |
| Training Requirement | No offline training needed | Requires extensive reward-labeled simulation episodes | Requires labeled historical dataset |
| Adaptability to new attacks | High (model-based; no retraining needed) | Moderate to High (if retrained periodically) | Limited (performance may degrade under unseen attacks) |
| Computational Resources Required | High: Real-time simulation and multi-node coordination | Medium: Lightweight neural policy inference | Low: Simple feedforward prediction |

IV. CONCLUSION

This paper proposed a novel Snitch Digital Twin framework for anomaly detection in Smart Voltage Source Converters of wind generator-integrated power grids. Unlike conventional learning-based models, Snitch-DT leverages a distributed virtual-physical system mirroring approach, enabling each VSC to locally validate its own operation while also participating in system-wide cross-verification.

Simulation results demonstrated the superior performance of the proposed method across various cyberattack scenarios. Compared to supervised ANN and DRL-based PPO detection, the Snitch-DT consistently achieved the highest detection accuracy, and F1 score with the lowest false positive and false negative rates. It also showed the fastest detection delay outperforming DRL and ANN.

Interestingly, although Snitch-DT was designed for detection, its early detection capabilities led to improved post-attack stabilization. The reactive power tracking RMSE was lowest with Snitch-DT, followed by DRL and ANN, indicating indirect mitigation benefits due to faster response. Thus, the Snitch-DT architecture presents a robust, low-latency, and high-accuracy solution and its ability to combine physical measurements with virtual replication makes it a promising candidate for next-generation cyber-physical grid security systems.

Future research will explore integration of Snitch-DT with mitigation mechanisms for dynamic setpoint correction, as well as adaptive digital twin models capable of online learning for improved robustness.

REFERENCES

[1] U.S. Department of Energy, "The Long-Term Strategy of the United States: Pathways to Net-Zero Greenhouse Gas Emissions by 2050," 2021.
[2] Y. Li and J. Yan, "Cybersecurity of Smart Inverters in the Smart Grid: A Survey," *IEEE Trans Power Electron*, vol. 38, no. 2, pp. 2364–2383, 2023, doi: 10.1109/TPEL.2022.3206239.
[3] "IEEE Std 1547.2-2023, IEEE Application Guide for IEEE Std 1547™-2018, IEEE Standard for Interconnection and Interoperability of Distributed Energy Resources with Associated Electric Power Systems Interfaces. IEEE Standards Coordinating Committee 21, 2023," 2023.
[4] "IEEE Std 2800-2022, 'IEEE Standard for Interconnection and Interoperability of Inverter-Based Resources (IBRs) Interconnecting with Associated Transmission Electric Power Systems,' IEEE Standards Coordinating Committee 21, New York, NY, USA, 2022.," *Electric Power Systems Research*, 2022.
[5] Z. Siahaan, E. Mallada, and S. Geng, "Decentralized Stability Criteria for Grid-Forming Control in Inverter-Based Power Systems," in *2024 IEEE Power & Energy Society General Meeting (PESGM)*, 2024, pp. 1–5. doi: 10.1109/PESGM51994.2024.10689037.
[6] U.S. DOE, "Electric Emergency Incident and Disturbance Reports," 2023.
[7] J. Shi, J. Wang, C. Wang, and Y. Liu, "LSTM-Based False Data Detection in Smart Grid," *IEEE Access*, vol. 7, pp. 134846–134858, 2019.
[8] M. Abdel-Basset, A. Mohamed, and M. A. Elhoseny, "An LSTM-Based Intrusion Detection System for Internet of Energy Networks," *IEEE Trans Industr Inform*, vol. 16, no. 12, pp. 8174–8185, 2020.
[9] R. Manandhar, M. Abdelrazek, and D. Zuehlke, "Machine Learning-Based FDI Attack Mitigation in Smart Inverters," *IEEE Trans Industr Inform*, vol. 18, no. 3, pp. 2010–2020, 2022.
[10] Y. Du, J. Qi, and M. Kezunovic, "Deep Reinforcement Learning for Online Cyber-Attack Detection in Smart Grids," *IEEE Trans Smart Grid*, vol. 12, no. 1, pp. 245–257, 2021.
[11] K. G. Vamvoudakis and J. P. Hespanha, "PPO-Based Cyber-Defense of Power Systems," in *Proceedings of the IEEE Conference on Decision and Control (CDC)*, 2020, pp. 6441–6446.
[12] B. Ahn, G. Bere, S. Ahmad, J. Choi, T. Kim, and S. Park, "Blockchain-enabled security module for transforming conventional inverters toward firmware security-enhanced smart inverters," in *2021 IEEE Energy Conversion Congress and Exposition (ECCE)*, IEEE, 2021, pp. 1307–1312.
[13] Danish Saleem *et al.*, "Modular Security Apparatus for Managing Distributed Cryptography for Command-and-Control Messages on Operational Technology Networks (Module-OT)," Jan. 2022.
[14] W. Danilczyk, Y. L. Sun, and H. He, "Smart Grid Anomaly Detection using a Deep Learning Digital Twin," in *2020 52nd North American Power Symposium (NAPS)*, 2021, pp. 1–6. doi: 10.1109/NAPS50074.2021.9449682.
[15] H. Pan, Z. Dou, Y. Cai, W. Li, X. Lei, and D. Han, "Digital Twin and Its Application in Power System," in *2020 5th International Conference on Power and Renewable Energy (ICPRE)*, 2020, pp. 21–26. doi: 10.1109/ICPRE51194.2020.9233278.
[16] M. A. H. Sadi, D. Zhao, T. Hong, and Mohd. H. Ali, "Time Sequence Machine Learning-Based Data Intrusion Detection for Smart Voltage Source Converter-Enabled Power Grid," *IEEE Syst J*, pp. 1–12, 2022, doi: 10.1109/JSYST.2022.3186619.
[17] M. A. H. Sadi, T. Hong, and M. H. Ali, "Deep Reinforcement Learning-Based Cyberattack Mitigation for Smart Voltage Source Converter-Enabled Power Grid," in *2024 IEEE Power & Energy Society Innovative Smart Grid Technologies Conference (ISGT)*, IEEE, 2024, pp. 1–5.
[18] B. Wu, Y. Lang, N. Zargari, and S. Kouro, *Power conversion and control of wind energy systems*. John Wiley & Sons, 2011.
[19] Y. Lu, M. Zhang, L. Nordström, and Q. Xu, "Digital Twin-Based Cyber-Attack Detection and Mitigation for DC Microgrids," *IEEE Trans Smart Grid*, 2024.